They each must
\documentstyle[aps,preprint]{revtex}
\def\today{\ifcase\month\or
           January\or February\or March\or April\or May\or June\or
           July\or August\or September\or October\or November\or
           December\fi
           \space\number\day, \number\year}
\begin{document}
\draft
\preprint{\vbox{Submitted to Phys.\ Rev.\ Lett.\hfill DOE/ER/40762--019\\
                \null\hfill UMPP \#94--039}}
\title
{On the $A$ dependence of $R = \sigma_L/\sigma_T$ and the\\
$Q^2$ dependence of Shadowing}
\author{J. Milana}
\address
{Department of Physics, University of Maryland\\
College Park, Maryland 20742}
\date{September 27, 1993}
\maketitle
\begin{abstract}
A higher--twist nuclear enhancement of $R = \sigma_L/\sigma_T$, as
might be expected
to arise due to fermi motion and whose magnitude is within the error-bars of
recent
experiment, is shown to lead to a monotonic {\it decrease} in the ratio
of nuclear vs. nucleon
cross--sections at small Bjorken $x$ for {\it increasing} $Q^2$.  This
effect at small $x$,
comparable in magnitude to those reported for shadowing, is driven
mainly by kinematic
factors and essentially vanishes for $x > .1$.  Its unusual $Q^2$
dependence rather complicates
the unravelling from present data in the shadowing region the
corresponding dependence in $Q^2$ of the nuclear structure functions,
$F_2^{A}(x,Q^2)$.
\end{abstract}
\newpage
\narrowtext
Shadowing, the observed depletion for heavy nucleii in the nuclear vs. nucleon
cross--section at low Bjorken $x$ in real or virtual photon scattering,
is a subject
of both continuing experimental and theoretical interest.  The
experimental situation
can roughly be seperated into two categories: those at low (or
zero)\cite{shadD_L} photon
virtuality, $Q^2$, and those at moderate to large\cite{shadD_H},
\cite{shadQ2} values
of $Q^2$.  It is the data at large $Q^2$ that is of concern for this
study, for which
Ref. \cite{shadQ2} is of particular interest as it is there that the
double--differential
cross--section $d^2 \sigma /dx dQ^2$ is presented for a range of $Q^2$ for a
number of $x$ values in the shadowing region.  What is to be here addressed is
our
ability to infer from this data the correct
$Q^2$ dependence of the structure functions $F_2^{A}(x,Q^2)$ in the shadowing
region.  This is important not only for hopefully distinguishing and
potentially selecting
different theoretical models\cite{shadTH1}, \cite{shadTH2} for
shadowing at large $Q^2$,
but because it is also crucial for our understanding the results of
other experiments such as
Drell--Yan\cite{DY} or $J/\psi$ production\cite{charm}, in which other
mechanisms\cite{Sean}, \cite{enemy} might be playing an important role.  It
should
be emphasized at the outset that this is not an attempt to explain shadowing,
for
which a diverse set of models presently exist and which can be roughly
seperated into
those using quark and gluon\cite{shadTH1} and those utilizing
hadronic\cite{shadTH2} degrees
of freedom.  Indeed, the effects here discussed all but vanishes for
the smallest $Q^2$ values
of Ref. \cite{shadQ2}.  Instead it is the point of this paper to
demonstate how a very simple,
higher--twist nuclear effect ({\it i.e.} which vanishes with increasing
$Q^2$) leads to a
rather unusual $Q^2$ dependence in the double--differential
cross--section, and one which
would, if verified, greatly complicate our ability to infer
$F_2^{A}(x,Q^2)$ from existing data.

To order $\alpha_e^2$, the double differential cross--section per
nucleon for lepton scattering
from an unpolarized nuclear target may be written as:
\begin{equation}
\frac{d^2\sigma}{dx dQ^2} = \frac{4 \pi \alpha_e^2}{Q^4} \frac{F_2(x,Q^2)}{x}
\left[ 1 - \frac{Q^2}{2MEx} -\frac{Q^2}{4E^2}+ \frac{Q^4}{8M^2E^2x^2}
\left(\frac{1 + 4 M^2 x^2/Q^2}{1 + R(x,Q^2)}\right) \right], \label{diff}
\end{equation}
where $M=.94$GeV is the mass of a (free) nucleon, $E$ is the incident
lepton's energy,
$F_2(x,Q^2)$ is the nucleon's structure function (in a nuclear medium), and
$R=\sigma_L/\sigma_T$ is the ratio of longitudinal to transverse
cross--sections,
which in terms of the usual functions $F_1(x,Q^2)$ and $F_2(x,Q^2)$ is given by
$R(x,Q^2)=F_L(x,Q^2)/2 x F_1(x,Q^2)$, where \mbox{%
$F_L(x,Q^2)=\left(1 + \frac{4x^2M^2}{Q^2}\right) F_2(x,Q^2) - 2 x F_1(x,Q^2)$}
is the so--called longitudinal structure function.  It is well known
that upto target mass
corrections, $R$ vanishes in the parton model for the case of spin
$1/2$ partons due to
helicity conservation.
In perturbative QCD, $R$ acquires a non--zero value from
$\alpha_s(Q^2)$ corrections to
the structure functions.  In addition to these corrections, $R$ is also
non--zero due to an
explicit higher--twist correction\cite{Rhtwist1}, \cite{Rhtwist2} to $F_L(x,
Q^2)$:
\begin{equation}
F_L^{\tau=4} (x,Q^2) = \frac{4 \Lambda^2}{Q^2} T_1(x),
\end{equation}
in which $T_1(x)$ is a new nucleon matrix element.  It involves
transverse components
of the covariant derivative acting upon quark field operators.  If one
ignores the fact
that this is a covariant derivative (required for gauge--invariance),
this new matrix
element corresponds naturally to what one associates with an intrinsic
transverse momentum
of the quarks in the parton model.  That is, one obtains in the limit that the
gluon
coupling in the covariant derivative $g \rightarrow 0$:\cite{Rhtwist2}
\begin{equation}
\lim_{g \rightarrow 0} \frac{4\Lambda^2 T_1(x)}{Q^2} = \frac{4}{Q^2}
\int d^2k_\perp k^2_\perp
f(x,k^2_\perp) = 4 \frac{\left \langle  k^2_\perp  \right
\rangle}{Q^2},\label{avgperp}
\end{equation}
where in the parton model $f(x,k^2_\perp)$ is related to $F_2(x)$ by
\begin{equation}
F_2^{\tau=2}(x) = x \int d^2k_\perp  f(x,k^2_\perp).
\end{equation}
Recent theoretical fits\cite{RfitTH} to $R$ for the free nucleon
indicate that such a
higher--twist element is indeed required to fit the data.

If we now turn to nuclear targets, it is clear form Eq. (\ref{diff})
that a nuclear dependence to $R$
could distort at any $x$ or $Q^2$ the extraction of $F_2^A(x,Q^2)$.  Indeed,
such a
conjectured dependence\cite{Rconj} was in fact proposed to account for
the initial discrepancies
between the EMC\cite{EMC} and SLAC\cite{SLAC} data for $x \approx .1$.  Besides
the ultimate experimental resolution of this discrepancy, the shift in
nucleii used by
\cite{Rconj},  $\Delta R \approx .15$, (taken to be essentially $Q^2$
independent, as would be
neccessary to fit the wide range of $Q^2$ available at these large
values of Bjorken $x$)
does not seem to be supported by later experiments\cite{Rmeas1} at SLAC
devoted to measuring
$\Delta R$.  We're now though interested in $\Delta R$ at small $x$,
for which Ref. \cite{Rmeas2}
is the sole source of experimental information.  This data (as was that from
\cite{Rmeas1}) is consistent with a value for $\Delta R = 0$, but with
such large error bars that
the best one should probably safely conclude is that $\Delta R < .1$.
However it should be
emphasized that because of the experimental difficulty in measuring
this quantity, a systematic
double--binning in both $x$ and $Q^2$ for $\Delta R(x,Q^2)$ in either
\cite{Rmeas1} or
\cite{Rmeas2} is lacking.

A simple reason to expect a nuclear dependence for $R$ is fermi motion.
 Returning to the
heuristics of the partonic model, motion of the nucleons in the nucleus
increases the average
transverse momentum of the quarks inside a nucleon.  By Eq.
(\ref{avgperp}), such an origin for
$\Delta R$ is higher--twist, and hence vanishes as $Q^2 \rightarrow
\infty$.  Despite this $Q^2$
dependence, it will be seen that the effect
on the measured cross-section $d^2 \sigma/dx dQ^2$ at given fixed $x$
actually grows with
$Q^2$.  This growth is a direct consequence of the kinematic factors
entering Eq. (\ref{diff}), as is
also the fact that for a given fixed $Q^2$, the effect grows for decreasing
$x$.

The world's data for $R$ for a free nucleon has been collected and
analysed by Ref. \cite{RfitE}
for which an empirical best fit has been obtained.  It is given by the formula:
\begin{equation}
R^{fit} = \frac{b_1}{{\rm ln} (Q^2/\Lambda^2)} \left( 1 + 12\frac{Q^2}{Q^2 + 1}
\frac{0.125^2}{0.125^2 + x^2} \right)+ \frac{b_2}{Q^2} + \frac{b_3}{Q^4 +
0.3^2}
\label{Rfiteq}
\end{equation}
where $\Lambda=.2$GeV, $\{b_i\}=\{.0635,.5747,-.3534\}$ and all momenta
are in GeV.
Note that there
is a misprint in \cite{RfitE} for the value of the first coefficent,
$b_1$: the decimal
point needs to be shifted one place to the left.%
\footnote{This misprint might account for the
apparent discrepancy between certain statements of Ref.
\protect{\cite{shadQ2}} and the results reported
here as to the relative importance that a nuclear dependence to $R$
generates in the differential
cross--section.}  Whereas the expression multiplying the inverse--log
term has been chosen to
fit the results from pQCD for $R$ at large $Q^2$, one should not
interpret too seriously the
various coefficents $\{b_i\}$ in terms of a twist expansion, as target
mass corrections must first be
disentangled.  Such a seperation was undertaken in Ref. \cite{RfitTH}.
For our purposes,
Eq. (\ref{Rfiteq}) is used to merely generate the best information we
presently have on $R$ for a
free nucleon.  One caveat is that no data actually exists for $R$ for
$x < .1$.  This fact will
be returned to later.

A higher--twist nuclear enhancement to $R^{fit}$ is conjectured.  To
$R^{fit}$ is added a term
$\Delta R$,
\begin{equation}
R^A = R^{fit} + \Delta R,
\end{equation}
modelled to fall as $1/Q^2$ {\it ala} Eq. (\ref{avgperp}) with $k_\perp =
.3$GeV
$\approx k_{fermi}$ taken.
No $x$ dependence is given this enhancement although such a dependence is
likely,
especially in the small $x$ region where the heuristic interpretation
of the higher twist matrix
element $T_1(x)$ is likely to break down\cite{Rhtwist2}.  In this
regard, the simple connection
with fermi motion is given merely as a plausibility arguement and the
skeptical reader could
interpret the following as simply an exercise exploring the effects
that introducing such a higher--twist
nuclear dependence for $\Delta R$ makes on the differential cross--section.

Fig. \ref{sigplot} plots the resulting ratio of nuclear to nucleon
cross--sections, Eq. \ref{diff},
such a $\Delta R$ generates for various values of Bjorken $x$ ($\{x_i\}
= \{.0125, .025, .05, .1\}$)
assuming no change in the nucleon to nuclear structure functions, {\it
i.e.} $F_2^A = F_2^N$.
The kinematics is that of Ref. \cite{shadQ2} ($E = 200$GeV) for which
the curves cutoff for
each $x_i$ at the kinematic upper limits of the experiment.  From top
to bottom the curves
correspond to decreasing $x_i$.  Note the trend with $Q^2$ for each
curve.  While the
plots in Fig. \ref{sigplot} are all consistently less than the
shadowing actually seen, especially at
small $Q^2$ (hence the initial statements that this is not an
explanation of shadowing),
for the larger $Q^2$ values the suppressions are large enough to play
an observable role
({\it e.g.} the shadowing for Calcium at $x=.025$ is roughly
$5\%$\cite{shadQ2}).
Observe most importantly that the effect at fixed $Q^2$ grows for decreasing
$x$.
This is dramatically demonstrated by the plot for $x=.0125$, where
though it is admittedly a
point for which both $R^{fit}$ and $\Delta R$ might be most seriously doubted.
It has been included nevertheless for emphasis.  In Fig. \ref{Rplot},
the solid line is the $x$ independent
 nuclear--enhancement, $\Delta R$, added to $R^{fit}(x,Q^2)$ of Eq.
(\ref{Rfiteq}).
For reference is also plotted $R^{fit}(x,Q^2)$ for two illustrative
values of Bjorken $x$.
The dashed line is $R^{fit}(x,Q^2)$ for $x=.1$ and the dotted line is
the (interpolated) value for $x=.025$.

The results in Fig. \ref{sigplot} are significantly more sensitive to
the assumed size of $\Delta R$
in the shadowing region than they are to $R$ itself.
For example, arbitrarily multiplying $R^{fit}$ by a factor of two makes
a change of only about
$.5 \%$ and $1 \%$ in the ratio of cross--sections for $x=.025$ and
$x=.0125$ respectively.
Since $R$ for the nucleon is  unknown in the shadowing region, this
insensitivity is relevant.
On the other hand, doubling (or halving) $\Delta R$ effectively doubles
(or halves) the effect at
large $Q^2$ for these same $x$ values.  As for the actual magnitude of
$\Delta R$ being introduced,
for $Q^2 > 5$GeV$^2$,
$\Delta R < .065$ and would appear to be small enough to fit within the
uncertainties (at any
Bjorken $x$) of experiment\cite{Rmeas1}, \cite{Rmeas2}.  Indeed, for
$Q^2=9$GeV$^2$, where the
effect in Fig. \ref{sigplot} for $x=.025$ is largest, $\Delta R < .04$,
well within experimental
error bars. For smaller values of $Q^2$ however, the $x$ independent
value of $\Delta R$ is in
apparent contradiction with the measured values\cite{Rmeas1} at $x >
.1$.  While this fact only
further emphasizes the result contained in Fig. \ref{sigplot} that a
nuclear dependence for $R$
yields negligible effects on the measured cross--section outside of the
shadowing region, theoretically
one notes that for the smaller $Q^2$ values, the magnitude of the
$\Delta R$ being introduced appears
less and less as a perturbation to $R(x,Q^2)$.  For example, at $Q^2 =
2$GeV$^2$, the used value
for $\Delta R = .18$ would be over a $60 \%$ correction to $R(x,Q^2)$
at $x=.1$.   In such
circumstances corrections of order $Q^{-4}$ cannot be justifiablely
ignored, and indeed a twist expansion
may not even converge\cite{comment2}.   For the future experiments at
CEBAF ($E=6$GeV) that will
measure $\Delta R(x,Q^2)$, this last consideration suggests that only
at the very highest $Q^2$ there
accessible might a twist--expansion in fact appear.  For of course the
main focus of this study,
the shadowing region, a machine with much higher energy (and presumably
an electron beam
for higher intensities in order to improve upon the CERN data, as in
{\it e.g} the proposed
$50$GeV SLAC upgrade) is neccessary for probing $\Delta R(x,Q^2)$.

In conclusion, the $Q^2$ dependence of Fig. \ref{sigplot} is rather
startling.  In a counter--intuitive
fashion, an intrinsically higher--twist effect for $\Delta R$ is of
growing importance with
increasing $Q^2$ in the observed rate. Examining Eq. (\ref{diff}) this
apparently surprising dependence
with $Q^2$ occurs because near the ends of phase--space, $1/xE$ enters
as an important new scale
where the parameter $y = Q^2/2MxE$ (the fractional energy loss of the electron)
is reaching its kinematic upper bound: $y \leq 1$.
In support of this effect, a close scrutiny of the data of Ref.
\cite{shadQ2} does indicate
an apparent systematic increase with $Q^2$ of the shadowing observed
for most of the
values of Bjorken $x < .1$.   While a nuclear enhancement of $R$ would
naturally account for such a
trend in the data, such an enhancement would also, depending on its
size for any particular $x$
in the shadowing region, act to mask the true $Q^2$ dependence of
$F_2^A(x,Q^2)$.  Without
therefore better knowledge of $\Delta R(x,Q^2)$ in the shadowing
region, the extrapolations of present
data to either yet higher $Q^2$ or other processes must be considered
problematic.

\acknowledgements  Much thanks to C. C. Chang, H. J. Lu and S. J. Wallace
for their many useful discussions, and to S. Rock for clarifying the
coefficents in $R^{fit}$.
This work was supported in part by the U.S. Department of Energy
under grant No. DE-FG02-93ER-40762.

\begin{figure}
\caption{ The ratio of nuclear to nucleon differential cross--sections
generated by a
$\Delta R \propto 1/Q^2$, as a function of $Q^2$ for given $x$.
{}From bottom to top, the curves correspond to
$\{x_i\}=\{.0125,.025,.05,.1\}$. \label{sigplot}}
\end{figure}
\begin{figure}
\caption{ In solid line, the value of the nuclear enhancement $\Delta
R$ being introduced.  For reference,
also the values of $R^{fit}(x=.1)$ in dashes, and $R^{fit}(x=.025)$ in
dots. \label{Rplot}}
\end{figure}

\begin{references}
\bibitem{comment2} There is little theoretical interest in cataloguing
the terms contributing to
order $Q^{-4}$ as the infrared divergences associated with these matrix
elements are explicitly
known not to factorize\protect{\cite{fact1}}.  Concerning factorization
for the $Q^{-2}$ terms, see
Ref. \protect{\cite{fact2}}.
\end{references}
\end{document}